\documentclass[conference]{IEEEtran}
\IEEEoverridecommandlockouts
\usepackage{cite}
\usepackage{pifont} 
\usepackage{amsmath,amssymb,amsfonts}
\usepackage{algorithmic}
\usepackage{graphicx}
\usepackage{textcomp}
\usepackage{xcolor}
\usepackage{tabularx}
\usepackage{array}
\usepackage{amssymb}
\usepackage{url} 
\newcolumntype{C}[1]{>{\centering\arraybackslash}p{#1}}
\def\BibTeX{{\rm B\kern-.05em{\sc i\kern-.025em b}\kern-.08em
    T\kern-.1667em\lower.7ex\hbox{E}\kern-.125emX}}
\begin{document}

\title{Strengthening DeFi Security: A Static Analysis Approach to Flash Loan Vulnerabilities\\
 }

\author{\IEEEauthorblockN{Ka Wai Wu}
\textit{Virginia Tech}\\
kawai@vt.edu \\
USA
}

\maketitle

\begin{abstract}
The rise of Decentralized Finance (DeFi) has brought novel financial opportunities but also exposed serious security vulnerabilities, with flash loans frequently exploited for price manipulation attacks. These attacks, leveraging the atomic nature of flash loans, allow malicious actors to manipulate DeFi protocol oracles and pricing mechanisms within a single transaction, causing substantial financial losses. Traditional smart contract analysis tools address some security risks but often struggle to detect the complex, inter-contract dependencies that make flash loan attacks challenging to identify.

In response, we introduce FlashDeFier, an advanced detection framework that enhances static taint analysis to target price manipulation vulnerabilities arising from flash loans. FlashDeFier expands the scope of taint sources and sinks, enabling comprehensive analysis of data flows across DeFi protocols. The framework constructs detailed inter-contract call graphs to capture sophisticated data flow patterns, significantly improving detection accuracy. Tested against a dataset of high-profile DeFi incidents, FlashDeFier identifies 76.4\% of price manipulation vulnerabilities, marking a 30\% improvement over DeFiTainter. These results highlight the importance of adaptive detection frameworks that evolve alongside DeFi threats, underscoring the need for hybrid approaches combining static, dynamic, and symbolic analysis methods for resilient DeFi security.
\end{abstract}


\section{Introduction}
Over the past few years, the finance industry has seen a tremendous growth in decentralized solutions with Decentralized Finance (DeFi) protocols due to the advantages offered by the underlying blockchain architecture. 
DeFi projections predict the revenue to be approximately USD 17.8 billion by 2023 and the number of DeFi users to be 22.09 million by 2028 \cite{statista}. Blockchain-powered DeFi enables consumers to have trustless transactions of digital financial assets (cryptocurrencies and tokens) without relying on a central party like a traditional bank. Over time, the DeFi landscape has evolved into a massive network with integrated financial instruments and protocols such as decentralized exchanges (DEXs), lending and margin trading platforms, liquidity managers, yield farmers, tracking indexes etc \cite{defiprotocols}. As of December 2023, the Total Value Locked (TVL) by DeFi protocols is USD 52.61 billion \cite{defillama}. This high value of digital assets managed by DeFi protocols makes them a lucrative target for attacks. \cite{Qian2023}. In the past few years, there have been several DeFi hacks with total hacked value amounting to USD 5.7 billion \cite{defillamaHacks}. 

\textbf{Motivation -} The humongous losses due to DeFi exploits can create a general loss of trust in the feasibility of DeFi as an alternative to traditional financial services. Therefore, it is imperative to enhance DeFi security measures and improve attack detection frameworks to avoid unwanted consequences of security breaches in DeFi protocols. DeFi protocols can be attacked by exploiting vulnerabilities such as re-entrancy, frontrunning, rug-pull etc \cite{SokDeFi}. One such common vulnerability in DeFi protocols is price manipulation which is often exploited to conduct hacks via flash loans.

Flash loans are uncollaterized loans which execute in one atomic transaction on the blockchain. DeFi protocols rely on price oracles to adjust the asset prices according to off-chain market factors. Insecure oracles are often vulnerable to price manipulation, and can hence lead to hefty flash loan thefts \cite{immunebytes}. A notable example of a flash loan attack is on the bZx platform \cite{PeckShield2021Dec} where the attacker leveraged a price oracle dependency of bZx on other DeFi platforms (i.e., Uniswap and Kyber) in manipulating cryptoasset exchange rates, making net profit of USD 318k within a single atomic transaction. 

Although there has been significant research progress to detect and fix bugs in smart contract codes, there has been limited research to detect price manipulation vulnerabilities as attackers exploit logic and design of DeFi protocols and their dependency on price oracles to conduct flash loan attacks. However, using program analysis and verification techniques for smart contracts, vulnerabilities with cross-contract dependencies can be identified. Common program analysis techniques include static analysis, dynamic analysis, symbolic execution and fuzzing. Furthermore, data flow in programs can be tracked via taint analysis methods. 

\textbf{FlashDeFier:} This work presents FlashDeFier, a static taint analyzer for smart contracts to detect price manipulation vulnerabilities. It is an extension of the existing state-of-the-art tool, DeFiTainter \cite{defitainter} which employs static taint analysis on decompiled smart contracts for price manipulation detection.

We highlight the following contributions of our work:
\begin{enumerate}
    \item We build upon the DeFiTainter framework to perform static analysis of inter-contract call and data flow by borrowing concepts from existing static taint analysis methods. 
    \item We expand the set of taint sources and sinks after an extensive study of decompiled bytecode of contracts. 
    \item We analyze the call flow graph of the smart contract to identify the function signature needed to build inter-contract data flow graph to track the propagation of taint from source to sink. 
    \item We evaluate our results and observe a 30\% improvement in detection accuracy as compared to DeFiTainter. 
\end{enumerate}

\section{Background}

\subsection{Motivating Example}

On November 6, 2020, a significant hack targeted Cheese Bank, a decentralized autonomous digital bank on Ethereum. This attack led to a loss of \$3.3 million in USDC, USDT, and DAI, exploiting a vulnerability in Cheese Bank's method of measuring asset prices using an AMM-based oracle, specifically Uniswap. The attacker utilized a flash loan to manipulate the collateral price on Uniswap, thereby enabling a series of malicious borrowing operations from Cheese Bank \cite{PeckShield2021DecCheese}.

\begin{figure}[htbp]
  \centerline{
  \includegraphics[width=0.5\textwidth]{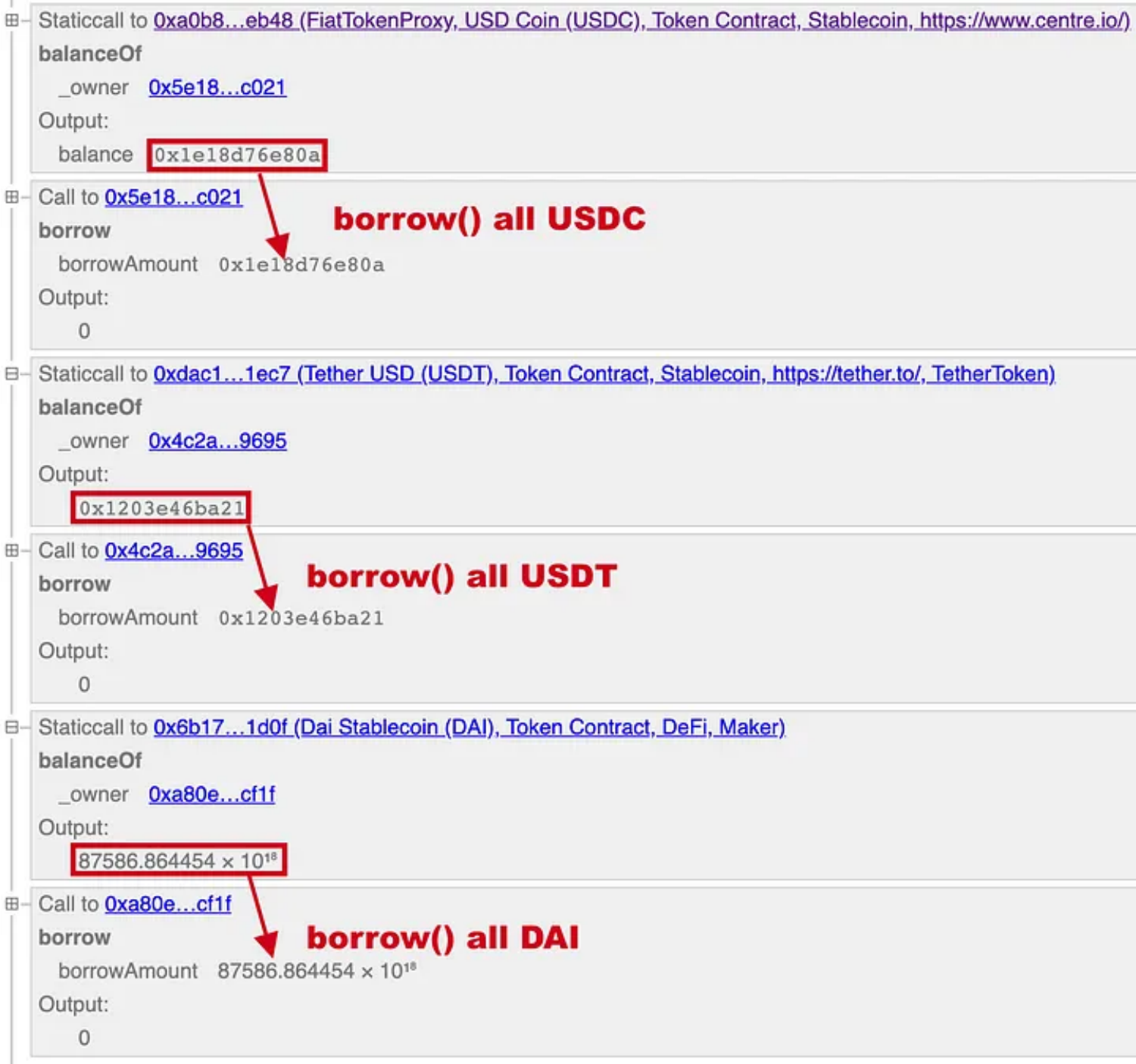}}
  \caption{Sequence of transactions the attacker used to check the balance of the liquidity pool of USDC/USDT/DAI before completely draining it from Cheese Bank on November 6, 2020.}
  \label{fig:motivating_example}
\end{figure}

The attacker began by taking a flashloan of 21,000 ETH from dYdX, then swapped 50 ETH for 107,000 CHEESE tokens on UniswapV2. Next, they added these CHEESE tokens and 78 ETH to the UniswapV2 liquidity pool, receiving UNI\_V2 LP tokens, which were then used to mint sUSD\_V2 tokens. These tokens served as collateral to borrow assets from Cheese Bank. The crucial part of the attack involved manipulating the CHEESE price by swapping 20,000 ETH for 288,000 CHEESE, significantly increasing the value of the UNI\_V2 LP tokens. The attacker then refreshed the price feeds in Cheese Bank to reflect this inflated value. With the manipulated prices, they drained 2 million USDC, 1.23 million USDT, and 87,000 DAI from Cheese Bank through legitimate borrow() calls (Figure \ref{fig:motivating_example}). The attacker then converted the CHEESE token back to approximately 19.98k ETH and 58,000 USDC to 132 ETH to cover the flashloan, collected the hacked assets into a single address, and returned the 21,000 ETH flashloan to dYdX. 

\subsection{Execution of Smart Contracts}
Smart contracts are stored as bytecodes on the blockchain and are executed by virtual machines embedded in the blockchain nodes. The most prevalent blockchain virtual machine today is the Ethereum Virtual Machine (EVM). Once the smart contract conditions are met and it is triggered, it is executed on the blockchain and this execution is broadcast as a transaction on the chain. The EVM translates the contract bytecode to a sequence of instructions and opcodes and execute the operations.  

\begin{figure}[htbp]
    \centering
    \includegraphics[width=0.98\columnwidth, trim={4.8cm 4.4cm 6.5cm 4cm},clip]{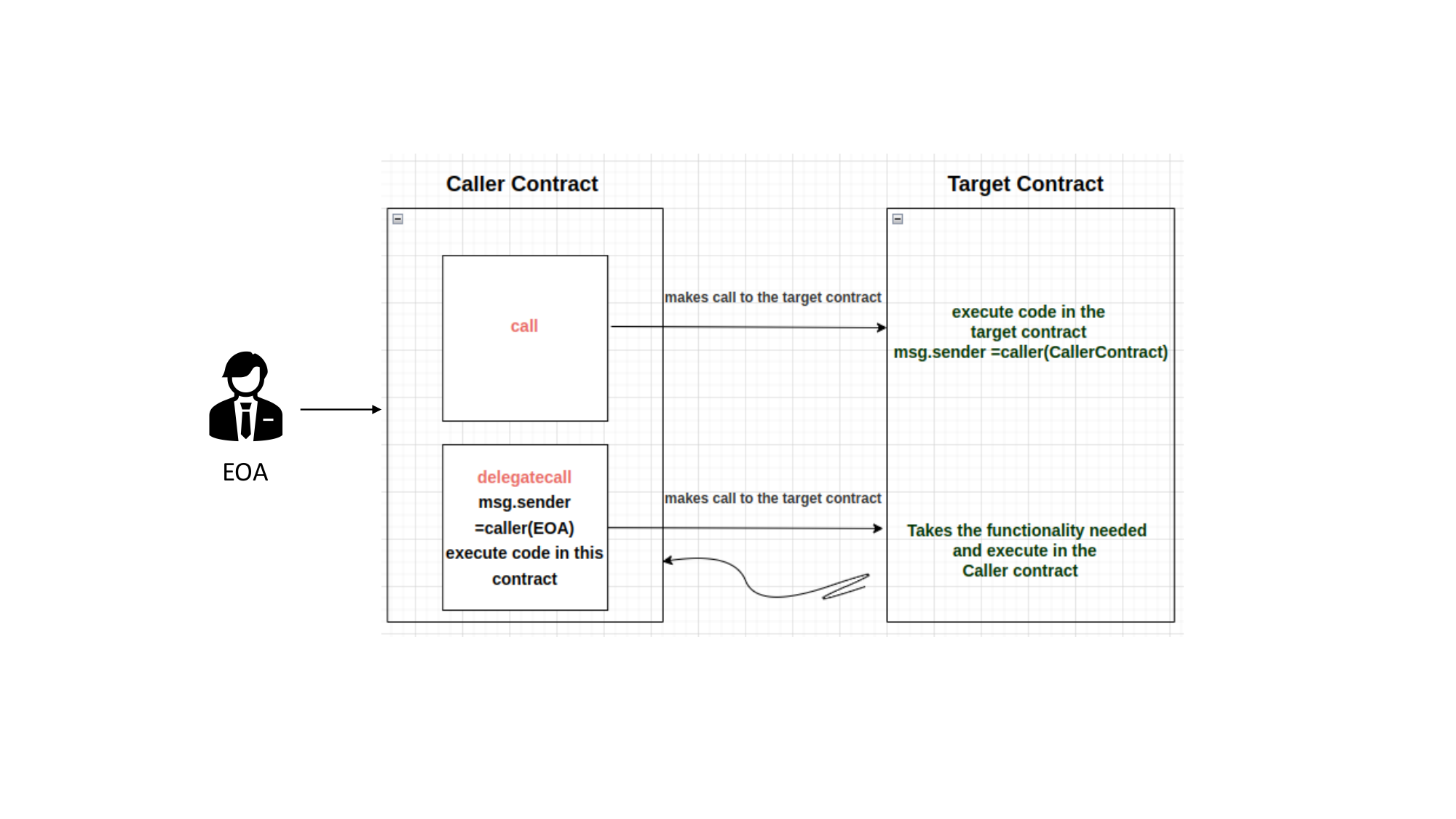}
     \caption{Difference between CALL and DELEGATECALL. Using DELEGATECALL, the execution is done in the storage of the calling contract i.e. the calling contract borrows code logic from another contract while retaining its identity.}
     \label{fig:opcodes}
\end{figure}

Ethereum smart contracts are commonly written in a high-level language called Solidity. An inter-contract vulnerability refers to those vulnerabilities within a call trace that involves more than two smart contracts. Such vulnerabilities are typically more complex to detect as it relies on the contextual information (i.e., contract attributes) during inter-contract invocation. A contract can call another contract deployed on Ethereum by referring to its address and it can be implemented by four opcodes: CALL, STATICCALL, DELEGATECALL, and CALLCODE. The CALLCODE method was deprecated since Solidity v0.5.0 in favor of DELEGATECALL \cite{solidityv0.5}. The STATICCALL method is similar to CALL except with the caveat that it doesn't allow modifications to the blockchain state. Figure \ref{fig:opcodes} shows the difference between CALL and DELEGATECALL.

\subsection{Flash Loans}
Flash loans are uncollateralized loans only valid within a single blockchain transaction. They allow traders to borrow assets without any collateral, under the condition that the borrowed amount is returned before the end of the transaction. This unique feature of flash loans is made possible by the atomicity of transactions on the blockchain, where a series of operations either all succeed or all fail together. If the loan is not repaid by the end of the transaction, the entire transaction is reversed, as if it never happened, ensuring that the lender does not lose their funds.

Adversaries can exploit the flash loan borrowing feature to borrow assets from one DeFi protocol, and then swap them on another such as an Automated Marker Maker (AMM) based DEX. An AMM is a type of DEX that allows users to buy and sell digital assets by algorithmically deciding the price of assets based on their liquidity in a given pool. The unique aspect of an AMM is its pricing mechanism, which is typically governed by a mathematical formula such as the constant product formula \( x * y = k \), where \( x \) and \( y \) represent the quantity of two different tokens in the liquidity pool, and \( k \) is a constant value. This formula ensures that the product of the quantities of the two tokens remains constant, thereby determining the price of the assets in the pool. As trades occur and the relative quantities of tokens in the pool change, the price of the tokens also adjusts automatically according to the formula.
significantly affecting the liquidity pool and thereby manipulating their market price. By injecting a large amount of one asset into a pool and removing another, an attacker can skew this ratio and thus the price. 

This kind of fluctuation in price of tokens in the liquidity pool is called price manipulation vulnerability and an attacker can exploit it via flash loans. One common method involves targeting thinly traded or low liquidity markets, where even small-scale trades can disproportionately sway market prices, thereby creating an illusion of significant market movement that doesn't accurately reflect true supply and demand dynamics. Additionally, manipulators often leverage information asymmetry, utilizing non-public information or disseminating misinformation to generate false market sentiments, thereby driving prices in a desired direction based on the reactions of uninformed traders.

\subsection{Taint Analysis}
Taint analysis is a technique employed in information flow oriented security and serves the purpose of identifying potential data flows from data with lower integrity (referred to as \textit{sources}) to data with higher integrity (known as \textit{sinks}). Taint sources are typically data that can be influenced or controlled by users, such as user-provided input data, while sinks are usually operations with security implications, like writing to a database. When data flows from sources to sinks, it indicates a potential security vulnerability. In context of smart contracts, this method can highlight areas where untrusted user inputs or external data sources may lead to unexpected behavior or compromise the integrity of the contract. 

\begin{figure}[hb!]
    \centering
    \includegraphics[width=0.98\columnwidth, trim={4cm 4cm 4cm 4cm},clip]{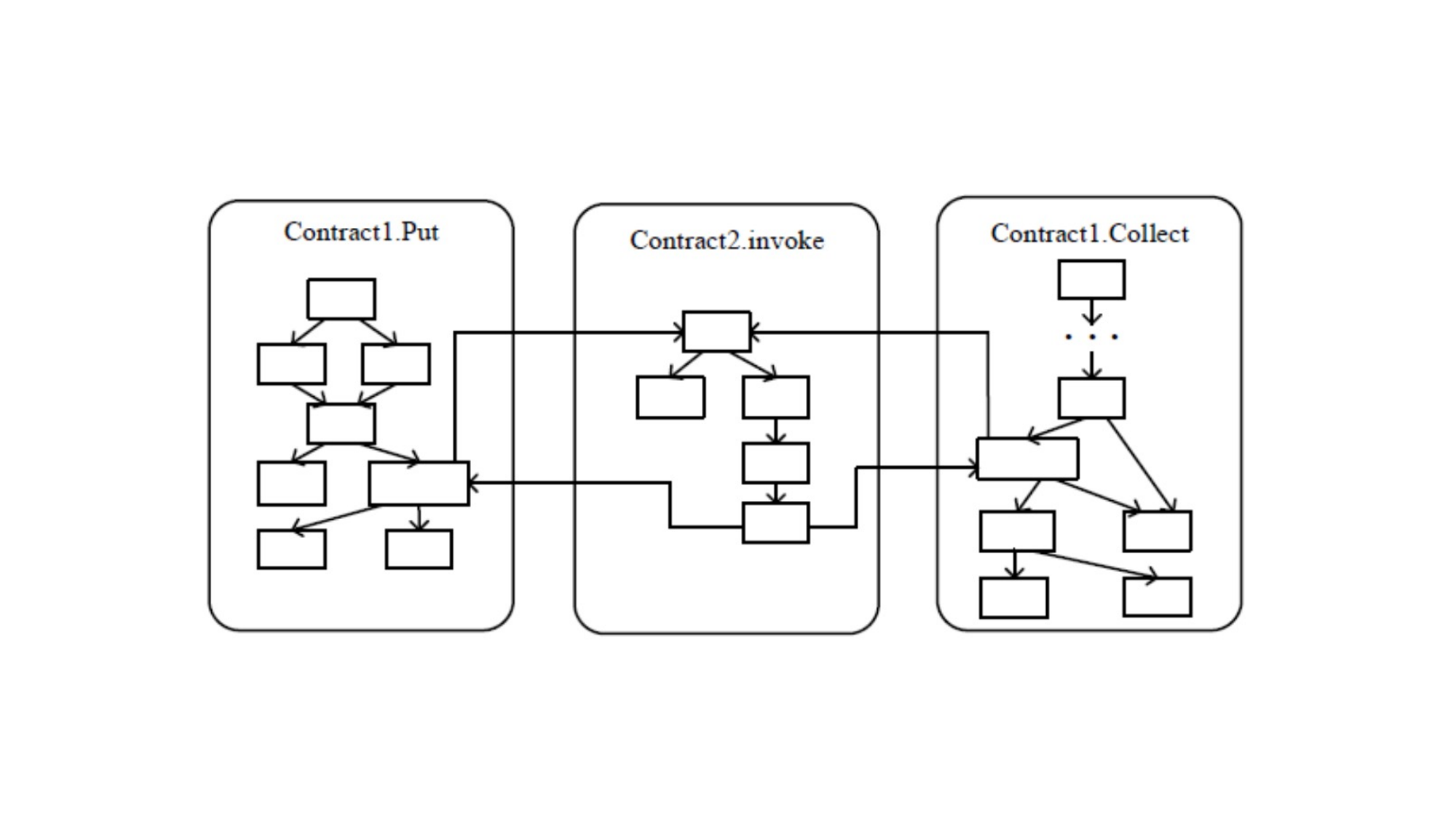}
     \caption{Call graph representation of function calls across contracts. This image shows how Contract 1 and Contract 2 interact with each other. The \textit{put} and \textit{collect} functions of Contract 1 calls \textit{invoke} function of Contract 2 which after some other calls returns the required value to function calls of Contract 1.}
     \label{fig:icpc}
\end{figure}

Smart contracts often interact with external contracts, thus, taint analysis should take into account the interactions and dependencies between contracts. In information flow analysis techniques, this type of inter-dependency can be represented by a call graph which captures function calls between different contracts (Figure \ref{fig:icpc}). The inter-contract call flow graph (ICCFG) is a directed graph where each node represents a function, and each edge represents a call between two functions. 

Taint analysis methods should be able to track the flow of tainted data in this complex inter-dependent graph. This can be achieved by specifying inference rules to track taint flow and needs the call graph to be comprehensive enough to not result in false negatives. Both static analysis or dynamic analysis methods to track taint flow. However, static analysis methods are simpler to execute than dynamic analysis which track data flow during program runtime.

\section{Related Works}
In this section, we highlight some notable research works in blockchain vulnerability. Traditional blockchain network security includes the impact on communication between nodes \cite{1,2,4,5, Wu2025Jan},  crypto solutions for privacy issues \cite{2230,2231} and machine learning based problems\cite{2232,2233,2234,2235}.  Recent studies on data-driven security analysis in decentralized networks \cite{sung2025housing, sung2025community, sung2025nlp} demonstrate the importance of leveraging machine learning, network structure analysis, and NLP techniques to enhance vulnerability detection. In section \ref{subsec:price}, we discuss some existing works for flash loan attack and price manipulation detection. Section \ref{subsec:taint} talks about some related works using taint analysis to detect vulnerabilties in smart contracts.  

\subsection{Price Manipulation Detection}\label{subsec:price}
\begin{table*}[t]
    \centering
    \caption{Comparison of Existing Price Manipulation Vulnerability Detection Tools}
    \label{tab:literature_comparison}
    \begin{tabularx}{\textwidth}{l c l X}
        \hline
        \textbf{Detection Tool} & \textbf{Open Source} & \textbf{Platform} & \textbf{Technique} \\
        \hline
        DeFiTainter \cite{defitainter} & \ding{51} & Ethereum, Binance, Polygon, Fantom & Static analysis of tainted data flow between contracts \\  
        DeFiScanner \cite{DeFiScanner} & \ding{55} & Ethereum & Deep learning-based feature extraction for attack detection \\
        DeFiRanger \cite{DeFiRanger} & \ding{55} & Ethereum & Cash Flow Tree (CFT) construction, semantic lifting, and pattern analysis \\ 
        DeFiPoser \cite{deFiPoser} & \ding{55} & Ethereum & Path pruning, parameter optimization, MDP-based strategy execution \\
        BlockEye \cite{blockeye} & \ding{55} & Ethereum & Symbolic analysis of oracle dependency and pattern-based runtime transaction validation \\ 
        ProMutator \cite{ProMutator} & \ding{51} & Ethereum & Grey-box fuzzing by mutating transactions and AMM data feeds \\ 
        FlashSyn \cite{flashsyn} & \ding{55} & Ethereum, Binance, Fantom & Counter-example driven polynomial approximation and interpolation \\
        \hline
    \end{tabularx}
\end{table*}
Several existing works have covered code vulnerability analysis in smart contracts and how to build attack detection tools for them. However, detecting price manipulation vulnerabilites in DeFi caused by flash loans is a logic vulnerability and thus more difficult to detect. We present a comparison of the existing works in detecting flash loan attacks leveraging price manipulation vulnerabilities in Table \ref{tab:literature_comparison}.

DeFiTainter \cite{defitainter} presents a static analysis tool to identify vulnerabilities using taint analysis. They construct an inter-contract call flow graph (ICCFG) and track the flow of tainted data from source to sink. DeFiRanger \cite{DeFiRanger} constructs a cash flow tree (CFT) based on transaction sequences, lifts the low-level semantics to a higher level and feeds them to an attack pattern analyzer to identify price manipulation vulnerabilities.  FlashSyn \cite{flashsyn} employs a counter-example driven numerical approximation and interpolation technique to generate transaction sequences which can potentially exploit price manipulation vulnerabilites.  However, FlashSyn being dependent on polynomial approximation is limited by the complexity of the smart contract involving many internal transactions to conduct flash loan attack. The authors of DeFiTainter compare their work with DeFiRanger and FlashSyn (which are dynamic analysis tools) and identify it as a superior detection tool.

DeFiScanner \cite{DeFiScanner} utilizes deep learning to extract event features from transactions in form of vectors, utilizes those vectors for high-level semantic generation and feeds them to a learning model to detect attack transactions. BlockEye \cite{blockeye} performs symbolic analysis on smart contracts to identify dependency on oracles and deploys a runtime transaction monitor to identify malicious patterns. DeFiPoser \cite{deFiPoser} employs SMT solvers to create profitable transactions in real-time on Ethereum blockchain. They discuss the adversarial strategy to quantify the threshold value for a MEV-aware miner to fork the blockchain by leveraging block-level state dependencies. ProMutator \cite{ProMutator} is a smart contract fuzzer which mutates transactions to assess price oracle attack susceptibility of a DeFi protocol. However, it does not simulate an actual flash loan attack and just mutates AMM data feeds for analysis. 

\subsection{Taint Analysis on Smart Contracts} \label{subsec:taint}
There are some existing studies which apply taint analysis to smart contract vulnerability detection. 
Ethainter \cite{ethainter} formulates an information flow analysis for smart contracts that takes into account the sanitization (i.e., guarding) coding practices in smart contracts to detect composite vulnerabilities. Pluto \cite{pluto} employs constructs an ICCFG to track the semantic information and supplements the ICCFG with dynamic transition information obtained from the runtime stack. It combines the taint analysis with symbolic path exploration with constraints to check path reachability. SmartDagger \cite{SmartDagger} recovers the contract attribute information from smart contract bytecode, and implements two graph optimization approaches to boost its analysis. Furthermore, eTainter \cite{etainter}  employs inter-contract taint analysis to identify gas-related vulnerabilities within smart contracts by analyzing the contract’s EVM bytecode.

\section{Design}
\begin{figure}[hb!]
    \centering
    \includegraphics[width=0.98\columnwidth, trim={0.1cm 5.6cm 0.3cm 6.2cm},clip]{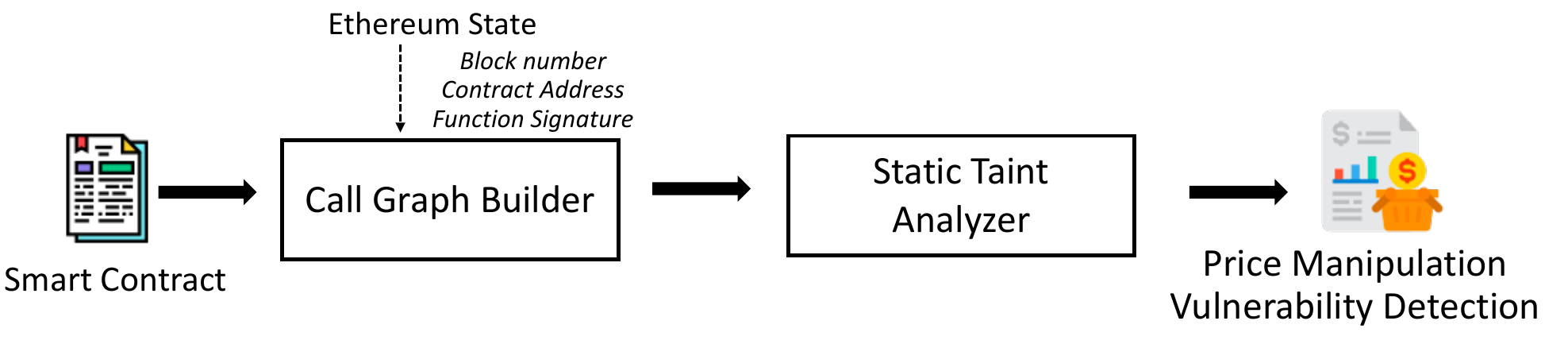}
     \caption{Design flow for detection of price manipulation vulnerability}
     \label{fig:workflow}
\end{figure}

After a thorough literature review of the existing works in the domain of price manipulation vulnerability detection, we decided to extend the work by DeFiTainter to enhance its detection accuracy. This choice was driven by the low complexity of static analysis methods and DeFiTainter already being the state-of-the-art as their comparison results show it to be better than other works like DeFiRanger and FlashSyn. We also chose to build our work upon DeFiTainter as their code is publicly available. Their method  constructing a call graph i.e. ICCFG and performing taint analysis on it for vulnerability detection (Figure \ref{fig:workflow}).

However, upon an extensive study and code review of their method, we identified some research gaps in DeFiTainer. Firstly, in their dataset, they include a function signature which acts as the root node of the ICCFG. However, their paper doesn't mention how this function is chosen and further manual code inspection makes us believe this choice to be arbitrary. It is important to select the correct function signature to build a more \textit{complete} ICCFG to track tainted data flow. Secondly, we believe the set of taint sources and sinks they have identified for the taint analyzer are insufficient. Since the flash loan attacks are getting more and more sophisticated, it is important to cover more taint sources and sinks to build a more \textit{complete} call graph which can reduce false negative results.  

Thus, our contribution in this work is to fill the aforementioned research gaps and build a more robust static taint analysis detection tool for price manipulation vulnerabilities.

\subsection{Call Information Restoration and Call Graph Generation}\label{subsec:cfg}

Each smart contract is stored in form of bytecode on the blockchain and has an associated persistent storage. Here, the bytecode is used to enable the program logic, and the persistent storage stores the state variables for contract attributes (e.g., identity, balance, message, address, etc.). DeFiTainter restores call information from the contract storage by providing a blockchain state, inducing the access location of the contract and extracting the stored content. Furthermore, DeFiTainter utilizes inter-contract data flow analysis to restore call information that originates from function parameters by passing on called function signatures. 

DeFiTainter uses a cyclical approach to alternate between call graph generation and inter-contract data flow analysis for call information restoration. However, they construct call graph only for a part of the execution path to prevent exploration of infeasible paths. The sub-call graph $SCG(x,y)$ consists of all calls directly or indirectly triggered by function $x$ of smart contract $y$ and is generated by tracking the calling relationship between functions. After generating $SCG(x,y)$, DeFiTainter performs taint analysis on $SCG(x,y)$ to detect price manipulation vulnerabilities in function $x$ of contract $y$. DeFiTainter repeats these two processes until a vulnerability is found or all functions of the detected DeFi protocol have been analyzed.

\subsection{Taint Analysis}
\begin{figure*}[ht!]
    \centering
    \includegraphics[width=0.8\textwidth, trim={1cm 7.8cm 1cm 0cm},clip]{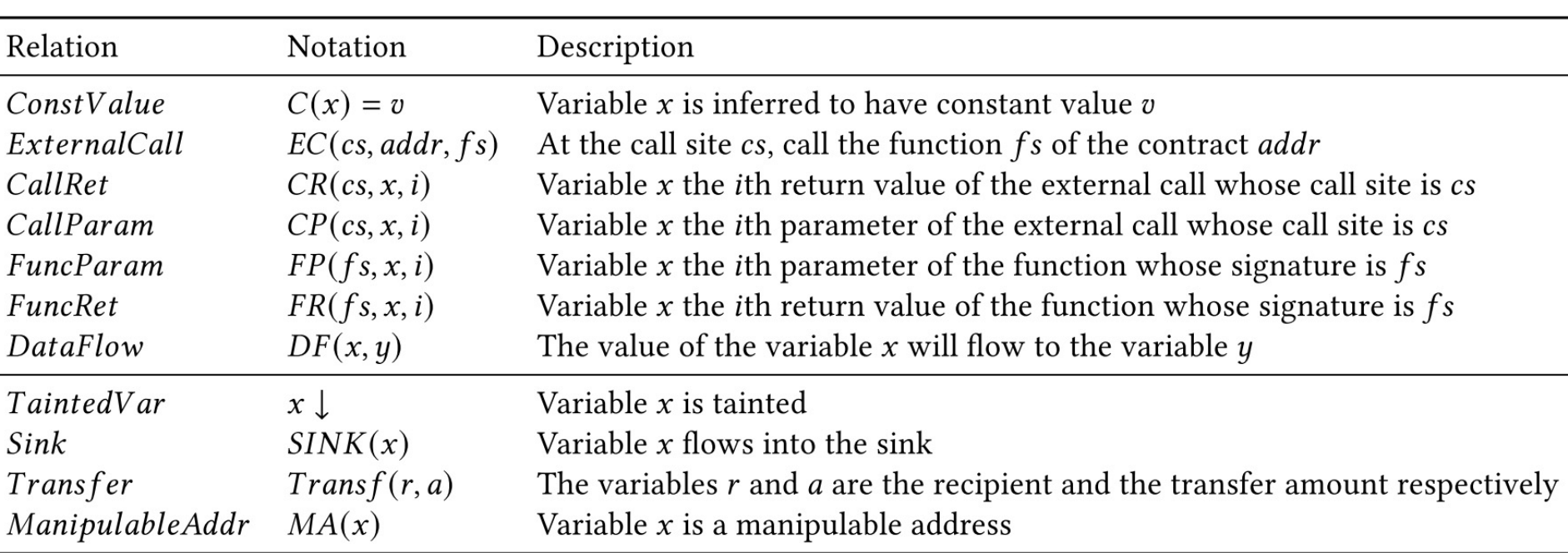}
     \caption{High-level semantic relations in DeFiTainter}
     \label{fig:semantic}
\end{figure*}

In their work, the authors of DeFiTainter construct an abstract input language for high level semantic extraction from smart contract. This induction of semantic relations is facilitated by Gigahorse \cite{gigahorse}, which is a state-of-the-art smart contract decompiler. Figure \ref{fig:semantic} illustrates the high-level semantic relations designed in DeFiTainter. 

\subsubsection{Taint Sources and Sinks}

\begin{figure}[hb!]
    \centering
    \includegraphics[width=\columnwidth, trim={5cm 3.9cm 6.6cm 0.3cm},clip]{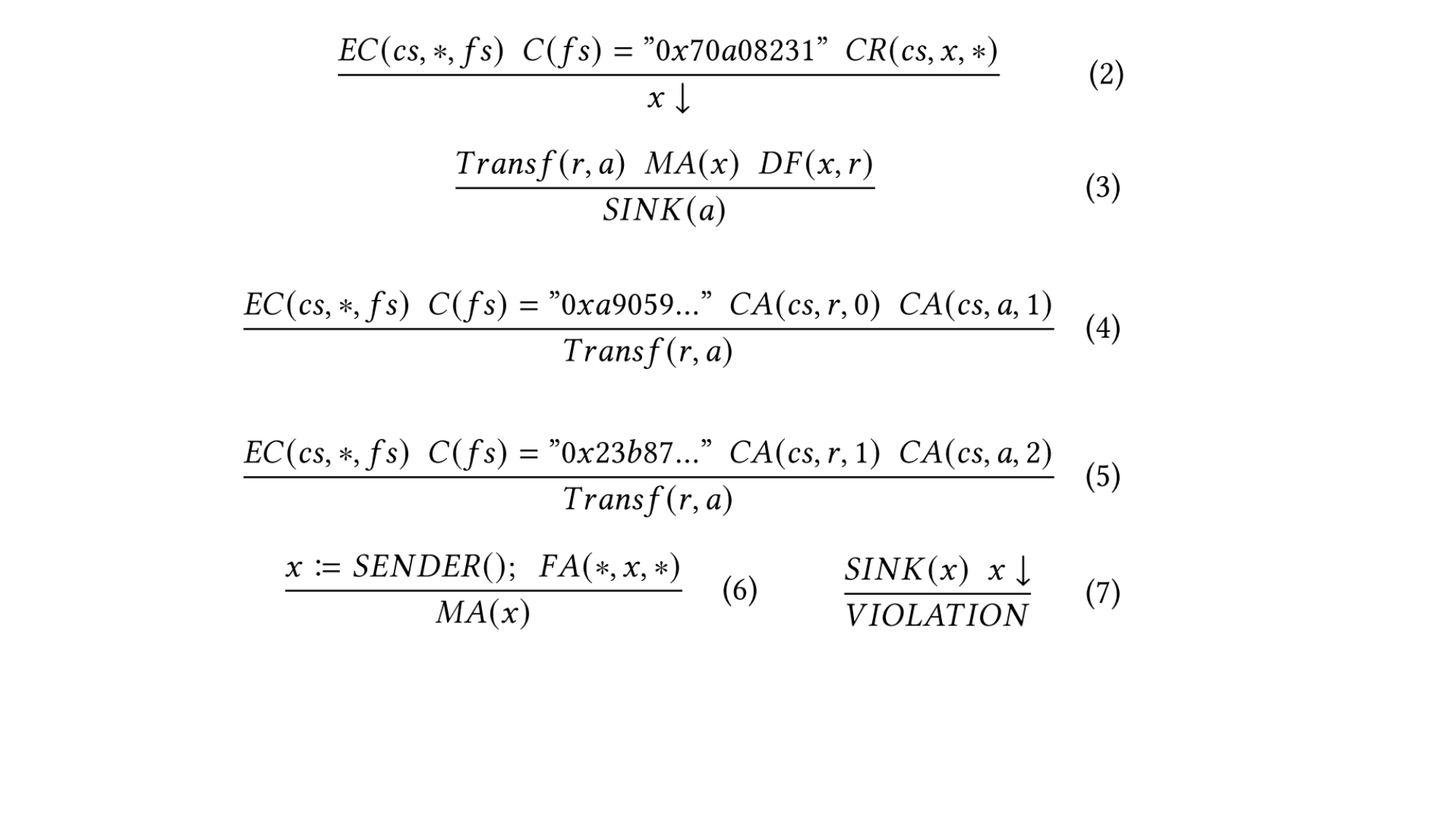}
     \caption{Inference rules defined in DeFiTainter for taint propagation based on $0x70a08231$ as function returning taint source and $0xa9059cbb$ and $0x23b872dd$ as sinks acquiring the taint from function input parameters.}
     \label{fig:rules}
\end{figure}

Identifying taint sources and sinks constitutes a fundamental pre-requisite for taint analysis and these taint labels vary across different vulnerabilities. For the price manipulation vulnerability, DeFiTainter marks the \textit{account balance} as a taint source as it can be manipulated by a flash loan. Furthermore, they mark operands of \textit{token transfer} operations as taint sinks since it can reflected the stolen amounts after price manipulation. They employ a manual marking of taint sources and sinks after an examination of attack events. They identify that account balance is usually checked with \textit{balanceOf(address)} function with a hex signature of $0x70a08231$, hence marking the return value of this function as tainted data. Furthermore, they identify that transfer operations in DeFi protocols are conducted through the \textit{transfer(address, uint256)} and \textit{transferFrom(address, address, uint256)} functions having signatures $0xa9059cbb$ and $0x23b872dd$ respectively marking the transfer amount received by these functions as the taint sink. The inference rules designed to conduct taint analysis in DeFiTainter are illustrated in Figure \ref{fig:rules}.

\subsection{Design enhancements}
Upon extensive study of attack events, we have identified the above definitions of taint sources and sinks to be insufficient. Therefore, we expand the set of taint sources and sinks in our work FlashDeFier, aiming to improve detection accuracy and reduce false negatives. These are mentioned in Table \ref{tab:TextSignature}. 

Other the the account balance, the return values of \textit{totalSupply(), getReserves()} can also be taint sources. This is because when conducting a flash loan attack affecting the liquidity of the DeFi protocol, the attacker usually tries to check the existing supply or reserves in the pool before manipulating their prices. The \textit{swap} function is usually called when an attacker swaps his tokens for the desired tokens after checking their liquidity. The \textit{approve} and \textit{allowance} functions are called when a sender allows another address (usually contract address) to transact funds on his behalf. This can be done by an attacker to make the tracing of the attack transaction more difficult by increasing complexity. 

Apart from the \textit{transfer} and \textit{transferFrom} functions, the stolen funds can also be propagated through functions such as \textit{buy, sell, withdraw}. Despite \textit{transfer} being the most commonly used functions, we add these functions to the taint sink set as they were observed in some attack scenarios.

\begin{table}[h]
   \centering
   \caption{Set of Function Signatures Associated with Taint Sources and Sinks in FlashDeFier}
   \label{tab:TextSignature}
   \begin{tabularx}{\columnwidth}{p{0.5\columnwidth} C{0.2\columnwidth} X}
    \hline 
    \textbf{Text Signature} & \textbf{Hex Signature} & \textbf{Taint Label} \\
    \hline
    balanceOf(address) & 0x70a08231 & Source \\  
    approve(address, uint256) & 0x095ea7b3 & Source \\  
    swap(address, int256, bool, uint160, bytes) & 0x24b31a0c & Source \\  
    allowance(address, address) & 0xdd62ed3e & Source \\ 
    totalSupply() & 0x18160ddd & Source \\ 
    swap(uint256, uint256, address, bytes) & 0x022c0d9f & Source \\
    getReserves() & 0x0902f1ac & Source \\ 
    transfer(address, uint256) & 0xa9059cbb & Sink \\ 
    transferFrom(address, address, uint256) & 0x23b872dd & Sink \\ 
    withdraw(uint256) & 0x2e1a7d4d & Sink \\ 
    buy(uint256, uint256) & 0xd6febde8 & Sink \\ 
    sell(uint256, uint256) & 0xd79875eb & Sink \\ 
    \hline
   \end{tabularx}
\end{table}

\section{Implementation}
To begin generation of $SCG(x,y)$ as mentioned in section \ref{subsec:cfg}, it is imperative to know function $x$ of contract $y$. We use existing transaction tracing tools like Etherscan \cite{etherscan} and BlockSec Phalcon Explorer \cite{phalcon} to trace the attack transaction. This enables us to find the logic address of the contract used to conduct the attack and the function signature which sits at the top of the ICCFG we want to construct. Figure \ref{fig:phalcon trace} shows the identification of contract address and function signature for the Upswing Finance attack in January 2023 \cite{upswingfinance}.

\begin{figure}[ht!]
    \centering
    \includegraphics[width=\columnwidth, trim={1cm 1cm 7cm 5.7cm},clip]{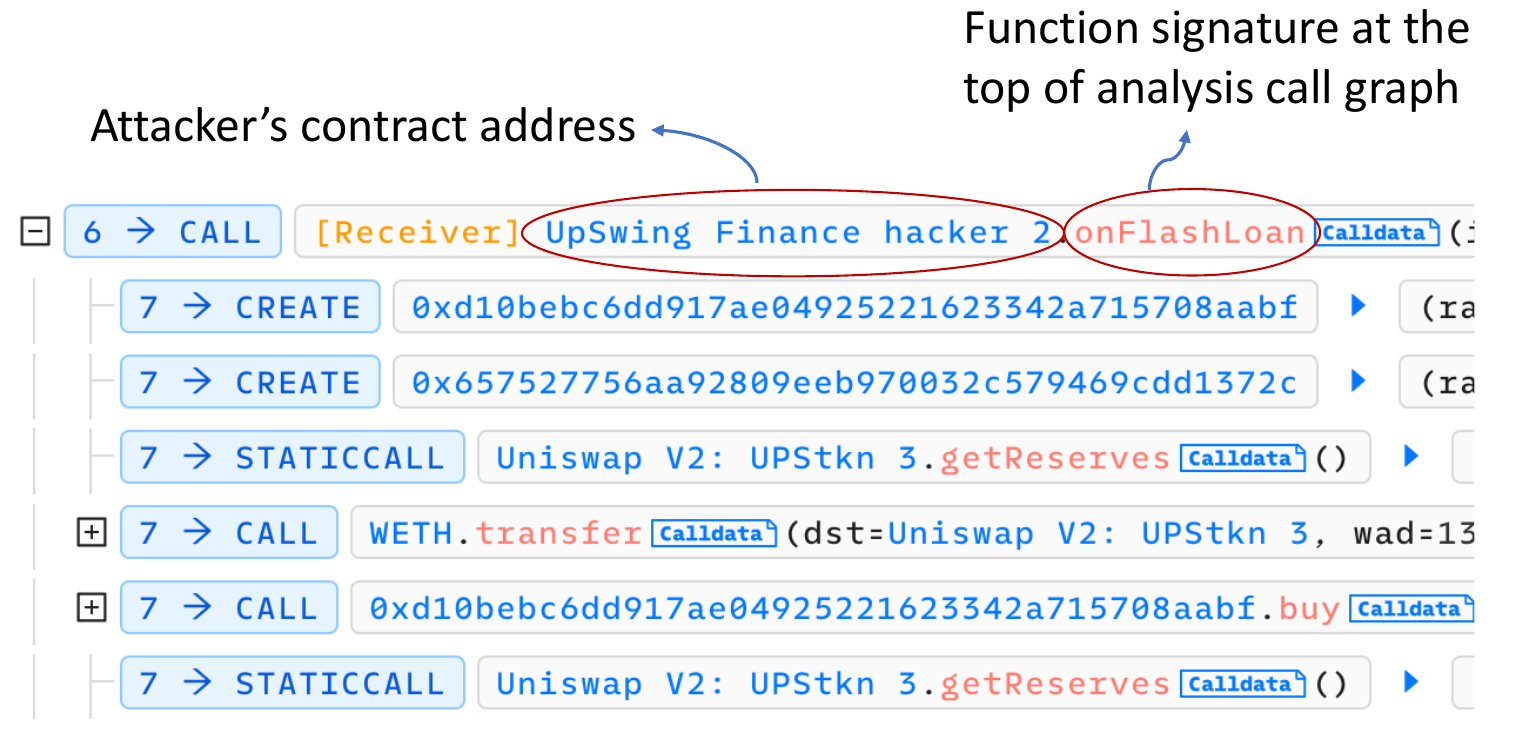}
     \caption{Upswing Finance Attack transaction trace from BlockSec Phalcon Explorer. As seen from highlighted values, the attacker's contract calls the function \textit{onFlashLoan} using the CALL method.}
     \label{fig:phalcon trace}
\end{figure}

\begin{figure}[ht!]
    \centering
    \includegraphics[width=\columnwidth, trim={2cm 9cm 3cm 1cm},clip]{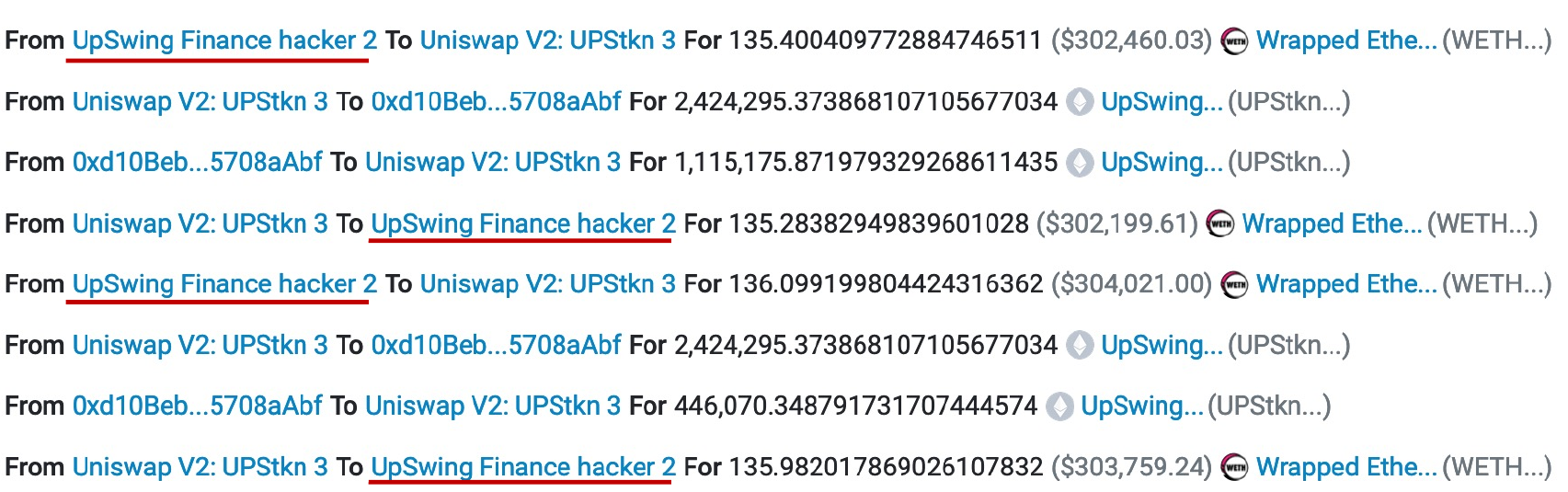}
     \caption{Upswing Finance Attack transaction trace from Etherscan. We can see the sequence of operations carried out by the attacker by invoking the malicious contract.}
     \label{fig:phalcon trace}
\end{figure}

\begin{figure}[ht!]
 \centering
 \includegraphics[width=\columnwidth]{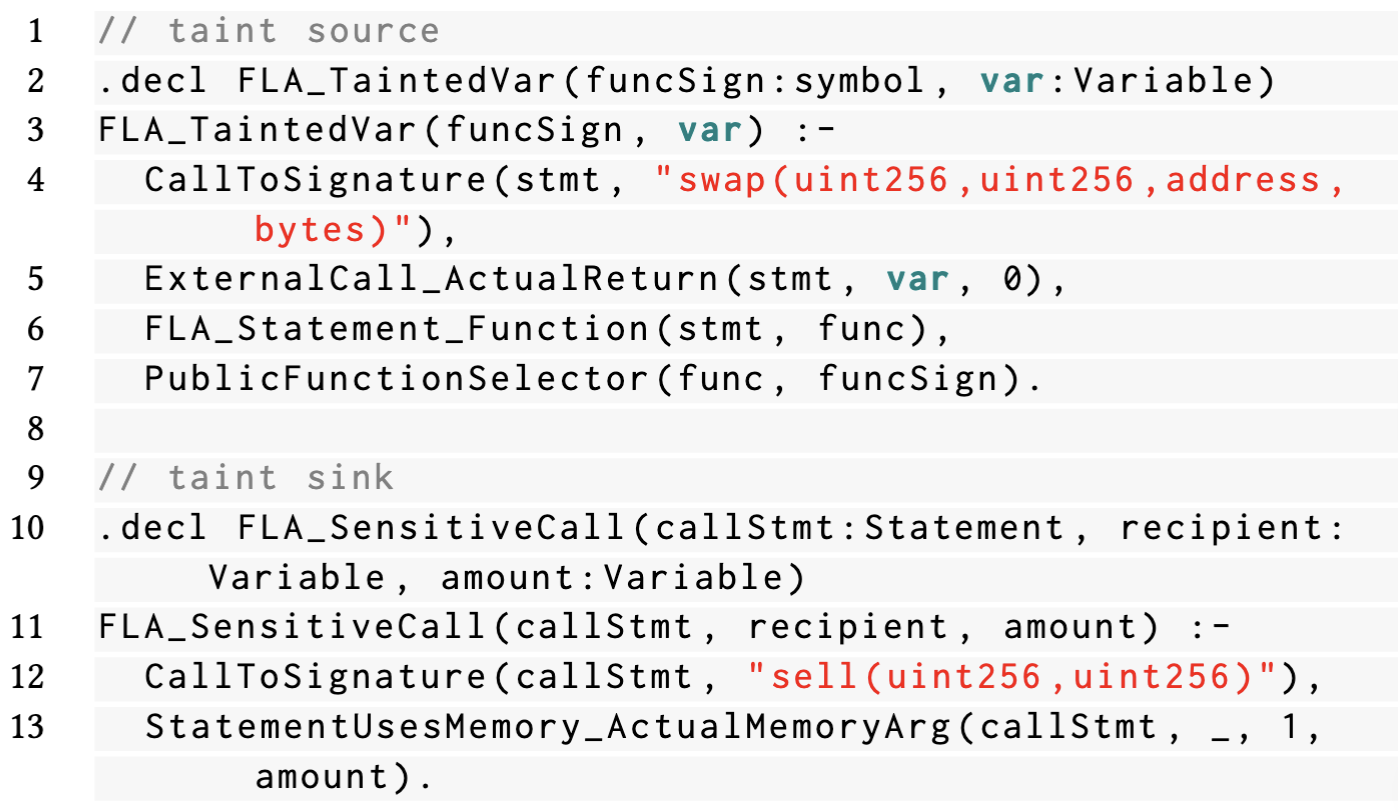}
\caption{Code snippet of FlashDeFier Datalog extensions on top of Gigahorse}
    \label{fig:dllcode}
\end{figure}

DeFiTainter is built on the Gigahorse framework that transfers EVM bytecode to its intermediate representation called GigahorseIR \cite{gigahorse}. Based on GigahorseIR, DeFiTainter extends a set of several hundred declarative rules in the Datalog language to restore call information and conducts inter-contract taint analysis for price manipulation vulnerabilities. Figure \ref{fig:dllcode} shows a snippet of the code modifications we have done to the Datalog client for Gigahorse. In addition, FlashDeFier obtains blockchain state information with Python web3 API \cite{web3api}, and connects the data flow between contracts with a python program.

We construct our dataset of attack transaction with field values such as exploit contract logic address, contract storage address, function signature of caller, block number on the Ethereum chain. This enables us to download the bytecode for the contract and analyzing it with our modified GigahorseIR to get a decompiled contract output.  

\section{Evaluation}
\subsection{Experiment Setup}
\textbf{Dataset.} We used a comprehensive dataset of historical attacks on the Ethereum mainnet to evaluate FlashDeFier. In particular, our dataset consists of high-value attacks that took place between 2021 and 2023 and is drawn from well-known DeFi rekt databases \cite{Web3rekt, DeFiREKTDatabase}. Web3rekt's data API was inaccessible, hence, we query these databases manually to identify which attacks were conducted via flash loans and more specifically using price manipulation as an extended method. Our investigation covers 17 attack incidents which caused a high loss of value and the same are mentioned in Table \ref{tab:DetectionResult}. The Ethereum Signature Database \cite{EthereumSignatureDatabase} serves as a repository for function and event signatures that are utilized within the Ethereum Virtual Machine (EVM). By querying this database, we were able to match the function text signatures with their respective hex signatures. 

\textbf{Environment.} We furnish an Ethereum endpoint on Quicknode \cite{quicknode} which allows interaction with the Geth client via JSON RPC API. Since, it allows only one node to be created in the free version, for the scope of this project, we limit our evaluations only to attacks on Ethereum mainnet. The experiments are performed on a computer running Ubuntu Desktop 22.04 on Intel i7-10700 CPU with 16 cores, each clocking 4.8GHz. 

\begin{table*}[h]
   \centering
    \caption{Detection Results}
    \label{tab:DetectionResult}
    \begin{tabularx}{\textwidth}{p{0.08\textwidth} p{0.13\textwidth} p{0.06\textwidth} X C{0.09\textwidth} C{0.08\textwidth} C{0.05\textwidth}}
    \hline 
    \textbf{Date} & \textbf{Incident} & \textbf{Loss (\$)} & \textbf{Exploited Contract} & \textbf{Function Signature} & \textbf{DeFiTainter} & \textbf{FlashDeFier} \\
    \hline
    29.09.2020 & Eminence Finance & 15M & 0x5ade7ae8660293f2ebfcefaba91d141d72d221e8 & 0xd79875eb & \ding{55} & \ding{51} \\  
    26.10.2020 & Harvest Finance & 33.8M & 0xf0358e8c3cd5fa238a29301d0bea3d63a17bedbe & 0xb6b55f25 & \ding{51} & \ding{51} \\
    6.11.2020 & Cheese Bank & 3.3M & 0x5e181bdde2fa8af7265cb3124735e9a13779c021 & 0xc5ebeaec & \ding{51} & \ding{51} \\
    14.11.2020 & Value.DeFi & 7M & 0x55bf8304c78ba6fe47fd251f37d7beb485f86d26 & 0x39bb96a8 & \ding{51} & \ding{51} \\  
    18.12.2020 & Warp Finance & 7.8M & 0xba539b9a5c2d412cb10e5770435f362094f9541c & 0xa555989 & \ding{51} & \ding{51} \\  
    21.07.2021 & Sanshu Inu & 280K & 0x35c674c288577df3e9b5dafef945795b741c7810 & 0x441a3e70 & \ding{51} & \ding{51} \\
    27.10.2021 & Cream Finance & 130M & 0x3d5bc3c8d13dcb8bf317092d84783c2697ae9258 & 0xda3d454c & \ding{55} & \ding{51} \\
    17.01.2023 & Upswing & 35K & 0xa3f47dcfc09d9aadb7ac6ecab257cf7283bfee26 & 0x22c0d9f & \ding{51} & \ding{51} \\
    26.01.2023 & Tom Inu & 35K & 0xb835752feb00c278484c464b697e03b03c53e11b & 0x22c0d9f & \ding{51} & \ding{51} \\
    09.03.2023 & SushiSwap & 28K & 0xc09707dc6917f098210568a80A88085D97F41bA3 & 0xf04f2707 & \ding{55} & \ding{51} \\
    15.04.2023 & 0x0.AI & 18K & 0x0f2b81e9d2771c5353346ada08e767eef3123ec9 & 0xf04f2707 & \ding{51} & \ding{51} \\
    17.04.2023 & DeFiGeek Japan & 20K & 0x0c845A1062F94d475c8303eCe4908cA2Bf98001f & 0xf04f2707 & \ding{55} & \ding{55} \\
    09.05.2023 & Weebcoin & 33K & 0xd8E1FED7B0238d1a31ee42BBec961d6bEd87C057 & 0xf04f2707 & \ding{51} & \ding{51} \\
    31.05.2023 & ERC20Token Bank & 111K & 0x7c28E0977F72c5D08D5e1Ac7D52a34db378282B3 & 0xf5c58cda & \ding{55} & \ding{55} \\
    14.09.2023 & Soda Finance & 84K & 0x1Bc1b75E99D941F6509FeFd70407aFB58329c2B4 & 0xf04f2707 & \ding{55} & \ding{55} \\
    23.11.2023 & KyberSwap & 47M & 0xaF2Acf3D4ab78e4c702256D214a3189A874CDC13 & 0x7b408b03 & \ding{55} & \ding{55} \\
    01.12.2023 & Fulcrum YFI & 208K & 0x03b7Bb750A974e0BD34795013F66B669f4110e54 & 0x10d1e85c & \ding{51} & \ding{51} \\
    \hline
    \end{tabularx}
\end{table*}

As it can be seen from Table \ref{tab:DetectionResult}, DeFiTainter is successfully able to detect 10 out of 17 incidents giving an accuracy of 58.8\%. In contrast, FlashDeFier is able to detect 13 out of 17 incidents, thus being accurate 76.4\% of the times. \textbf{Thus, FlashDeFier outperforms DeFitainter by 30\%.} It is worth noting the incidents which FlashDeFier fails to detect are also missed by DeFiTainter. In case of the KyberSwap attack, the attack was extremely sophisticated and targets the reinvestment curve feature with repeated swaps to manipulate liquidity tick range \cite{SlowMist2023Nov}. These kind of highly sophisticated attacks cannot be identified with simple static analysis techniques. 

Due to time constraints, we haven't done timing measurements to compare the detection time of FlashDeFier vs DeFiTainter. However, we expect the time taken for detection to be higher in FlashDeFier since it expands the set of taint sources and sinks, thus computing more paths in the call flow graph.  

\section{Discussion}
With the growth of DeFi, attackers are also getting more sophisticated. For instance, in the recent Kyberswap attack, the attacker drains the liquidity pool of Kyberswap which uses CLMM (concentrated liquidity market maker) to reduce slippage. CLMM is supposed to be better than AMM which have previously been the target of many flash loan attacks (e.g., Cheese Bank). Thus, we believe there is significant room for research in vulnerability detection and attack flow analysis of smart contracts, especially when it comes to flash loans and price manipulation since these exploit design flaws, making the attack harder to detect. The attack detection methods also need to be hybrid to leverage the benefits offered by each. For instance, while static analysis methods can be easier to compute and deploy on-chain because of their low complexity, they can lead to more false positives. Symbolic execution methods can give more precise solutions but can lead to path explosion problems. Hence, a combined solution of static analysis and symbolic execution can generate more comprehensive analysis with fewer false positives. One solution we can think of is using static analysis to limit the size of the graph and then performing symbolic execution with path constraints on it.

Furthermore, for taint analysis specifically, machine learning algorithms can be used to train on previous flash loan based price manipulation attacks to learn taint sources and sinks which can be fed to the smart contract decompiler. Lastly, most of the existing research work focuses their solutions on post-incident analysis. However, there is an urgent need of a robust attack detection tool which can be deployed on-chain to prevent the loss of funds as soon as suspicious activity is observed. 

Finally, there is a growing need for standardized security protocols and best practices in the DeFi space. Establishing industry-wide security standards and regular audits can help in identifying and mitigating vulnerabilities before they are exploited. Collaboration among DeFi platforms, security researchers, and regulatory bodies can facilitate the development of these standards, leading to a more secure and stable DeFi ecosystem.

\section{conclustion}
The focus of this paper is on the development and evaluation of advanced detection tools for identifying vulnerabilities in decentralized finance (DeFi) protocols, with a particular emphasis on those exploited through flash loans and price manipulation. We address the challenges and complexities involved in safeguarding DeFi protocols against such sophisticated attacks. FlashDeFier is a detection tool that extends the capabilities of the existing state-of-the-art tool, DeFiTainter, using static taint analysis on decompiled smart contracts. The evaluation of FlashDeFier demonstrates its efficacy, as it successfully detected 76.4\% of incidents in a comprehensive dataset of historical attacks on the Ethereum mainnet, marking a significant 30\% improvement over DeFiTainter. This achievement underscores the potential of incorporating advanced techniques in the realm of blockchain security. We also provide further discussion about the limitations of current methodologies in detecting highly complex attacks. We offer insights that future research should focus on enhancing detection methods and expanding the scope of analysis to cover a wider range of attack vectors, thereby strengthening the resilience of DeFi systems against evolving threats. This entails not only improving static analysis methods but also integrating dynamic and symbolic execution methods to form a more holistic and robust approach to vulnerability detection in the rapidly growing and evolving landscape of decentralized finance.

\end{document}